Comments on Paper by D. J. Antonio et al. **"Piezomagnetic switching and complex phase equilibria in uranium dioxide"** in *Communication Materials*, https://doi.org/10.1038/s43246-021-00121-6, published 5-Feb-2021.

**Comments on diffraction experiments at high-field on UO$_2$ in the ordered state**
G. H. Lander[1,2] and R. Caciuffo[1]
[1]*European Commission, JRC, Postfach 2340, D-76125 Karlsruhe, Germany*
[2]*Institut Laue-Langevin, Boîte Postale 156, F-38042 Grenoble, France*

In a recent diffraction study of a single crystal of UO$_2$ at low temperature (15 K) and high magnetic fields (up to 25 T) the authors [1] have observed that a diffraction peak from the (888) reflection splits into *two* peaks that move in opposite directions in 2-theta with increasing field. They are not able to understand the appearance of the two peaks, claiming the splitting is "unexpected". This experiment follows earlier work on the same material measuring the strain as a function of field, which suggested a phase transition when the field was applied along the <111> axis below the ordering temperature of 30 K [2, 3].

*What the authors have observed is that a sufficiently high magnetic field along the [111] direction with UO$_2$ below $T_N$ (= 30 K) induces a rhombohedral distortion of the unit cell.*

If such a rhombohedral distortion nucleates around each of the four different <111> directions (neglecting time reversal) in the single crystal, then the degeneracy of the *d*-spaces will be lifted and the *d*-space for the domain **A** in which **H** || [111], will be slightly different from the *d*-spaces in domains **B**, in which the remaining three <111> directions are at an angle of 70.53º to **H**.

Such a lifting of degeneracies is treated in many articles concerning rare-earth and actinide materials, in which the orbital-lattice interaction is strong. A convenient paper is from 1974 [4], where Table I gives the behavior of different sets of reflections for different types of distortions. For a rhombohedral system it is convenient to construct hexagonal unit cells, with the $c_H$ axes along the different <111> axes. As discussed in that paper [4] the cubic (HHH)-type of reflections will split into *two*, and only two, reflections with the hexagonal indices (H0.H) and (00.3H), with multiplicities *m* of 6 and 2, respectively. This is what they have observed in the experiments on UO$_2$.

Moreover, if the relative change in the $c_H$ axis is δ, where this quantity is small (see Ref. [1], Fig. 2) and is – 660 x 10$^{-6}$ for an applied filed of 20 T with the sample at 15 K, then the relative change in the $a_H$ axes is ½ that value, and must be in the opposite direction [5]. The relative change in $c_H$ can be deduced directly from the (00.3H) reflection, shown in red in Fig. 2 of the paper and referred to as domain **A**. However, the (H0.H)-type reflections, coming from domains **B**, (shown in blue in Fig. 2) will not have a variation given by δ/2, but simple algebra shows that the position of the peak should vary by δ/3. This is *exactly* what is observed in Fig. 2, where the slopes of the two lines (red and blue) differ by a factor of three, and are in opposite directions. The overall strain (at 20 T and 15 K) is readily determined to be ε = 3δ/2 = – 990 x 10$^{-6}$. This is comparable to the strains of other actinide samples as tabulated in Ref. [4]. Actinide ferromagnets show larger effects, as there is direct coupling of the moments to the lattice; in US, for example, the strain associated with the rhombohedral distortions at $T_C$ is an order of magnitude greater than found here in UO$_2$.

The authors of Ref. [1] could provide useful information by showing plots of the domain populations as a function of field cycles. Trivially, the domain **A** population is determined by Intensity domain **A**/(Int. **A** + Int. **B**). A random domain population would give 25% domain **A**. By eye from their plots, it would appear that this population may be increased to ~ 70% of domain **A** by training with the magnetic field. From their data (limited to H = 20 T) it appears that a single domain sample is not achieved, which is at variance with the assumption of a single domain in Ref. [2]. Perhaps a single domain can be achieved at much higher fields.

With the introduction of a finite rhombohedral distortion, and the resulting *intra*-domain stresses observed by the x-ray diffraction measurements [1], one must take account also of the *inter*-domain stresses. The crystal will clearly try to minimize the latter but fails, as it is the total sum of both these strains across the crystal that are observed in the bulk measurements reported earlier [2, 3]. However, the bulk strains are, on average, at least an order of magnitude smaller than the *intra*-domain strain caused by the distortion. *Inter*-domain stresses will influence the domain sizes (which could be determined by high-resolution diffraction), probably act against a single domain being achieved, and are partly responsible for the distortions appearing in the **B** domains (since the applied field is not along the local <111> axes in these domains), as well as the hysteresis and "butterfly" effects observed earlier [2]. The longitudinal strain in Ref. [2] reaches + 150 ppm (Fig. 2) with 18 T applied at 2.5 K, whereas the diffraction experiments (Ref. [1], Fig. 2, red results) show that with a single domain the longitudinal change in length would be – 660 ppm. The large discrepancy between these two numbers, and their opposite sign, reflects the importance of the *inter*-domain stresses.

In conclusion, our note fully explains the observation of *two* reflections and the dependence on increasing field [1]. It opens the way for a better understanding of the earlier bulk strain effects [2, 3], but it also allows speculation on the exact configuration of the magnetic moments. We know that an antiferromagnetic will try to align its opposing moments *perpendicular* to the applied field, which suggests that with the breaking of symmetry in the high-field state, the moments will be perpendicular to the <111> axes. One possibility is that the moments are aligned along <110> directions, which would imply that $UO_2$ at high field is a 2-***k*** configuration, rather than the 3-***k*** configuration known to exist at zero field below $T_N$ = 30 K [6]. This is speculation from the x-ray diffraction results in Ref. [1], but could be proved by neutron diffraction or resonant x-ray diffraction at the uranium *M* edges. Finally, there should be a re-ordering of the electric quadrupole moments known to order below $T_N$ in $UO_2$, as well as the displacements of the oxygen atoms associated with such quadrupolar ordering [7].

*lander@ill.fr,

31-March-2021

**Author Contributions**
Both authors have contributed equally to the manuscript

**Competing Interests**
Both authors declare no competing interests.